\documentclass{article}
\usepackage{spconf,amsmath,graphicx}
\usepackage{amsfonts}
\usepackage{amsmath,graphicx, verbatim}
\usepackage{xcolor}
\usepackage{multirow}
\usepackage{float}
\usepackage{makecell}
\usepackage{amssymb}
\usepackage{algorithm}  
\usepackage{algpseudocode}  
\usepackage{pgfplots, hyperref}
\usepackage[raggedrightboxes]{ragged2e}
\usepackage{cancel}
\usepackage{booktabs,multirow}
\usepackage{adjustbox}
\usepackage[font=small]{caption}
\usepackage{subcaption}
\usepackage{enumitem}
\usepackage{pifont}

\usepackage[
backend=biber,
style=ieee,
citestyle=numeric-comp,
maxbibnames=3,
maxcitenames=3,
doi=false,isbn=false,url=false,eprint=false
]{biblatex}
\defbibheading{bibliography}[\refname]{}

\title{Speech Recognition for Analysis of Police Radio Communication}
\name{Tejes Srivastava\textsuperscript{1}, Ju-Chieh Chou\textsuperscript{2}, Priyank Shroff\textsuperscript{1}, Karen Livescu\textsuperscript{2}, Christopher Graziul\textsuperscript{1*}\thanks{\textsuperscript{*}Corresponding author: \href{mailto:graziul@uchicago.edu}{graziul@uchicago.edu}}}
\address{$^1$University of Chicago, Chicago, IL \\
$^2$Toyota Technological Institute at Chicago, Chicago, IL}

\begin{document}
%
\maketitle
\begin{abstract}
Police departments around the world use two-way radio for coordination. These broadcast police communications (BPC) 
are a unique source of information about everyday police activity and emergency response. Yet BPC are not transcribed, and their naturalistic audio properties make automatic transcription challenging. We collect a corpus of roughly 62,000 manually transcribed radio transmissions (\textasciitilde46 hours of audio) to evaluate the feasibility of automatic speech recognition (ASR) using modern recognition models. We evaluate the performance of off-the-shelf speech recognizers, models fine-tuned on BPC data, and customized end-to-end models. We find that both human and machine transcription is challenging in this domain.  Large off-the-shelf ASR models perform poorly, but fine-tuned models can reach the approximate range of human performance.  Our work suggests directions for future work, including analysis of short utterances and potential miscommunication in police radio interactions.  We make our corpus and data annotation pipeline available to other researchers, to enable further research on recognition and analysis of police communication.
\end{abstract}
\section{Introduction}
\label{sec:intro}
In the last few years, there has been increasing interest in studying the language of police in the United States, due in part to public demand for better police accountability, especially surrounding questions of racial disparities in police encounters~\cite{hetey2024cruiser}. Recent work on the language of policing has focused on the speech recorded in body-worn police cameras during traffic stops~\cite{voigt2017language,fielddeveloping}.  This work has found racial disparities in such interactions~\cite{voigt2017language}, including differences that appear near the very beginning of police stops~\cite{rho2023escalated}. 
 
 However, the language used among policing professionals, outside of their direct interaction with community members, is less studied.  One source of such data is two-way radio transmissions involving police officers and dispatchers. In these broadcast police communications (BPC), dispatchers communicate with police officers, and officers communicate with each other, about police incidents and the locations and people involved.  The language used among policing professionals is a rich source of information about these events and may shed light on the preparation {\it leading up to} police interactions with the public. As with body worn cameras, recent work finds that racial disparities exist in BPC, where use of two-way radio poses unacknowledged privacy risks linked to disproportionate communication about members of specific groups (e.g., Black or African American males)~\cite{vargas_digital_2019,venkit2024race}. However, this work has only been able to analyze a small fraction of available BPC due to the resource intensive nature of manual transcription. 

Police radio transmissions are available in large quantities, but they are noisy and unannotated.  A prerequisite for expanding the study of the language of police radio, therefore, is to develop transcription and related annotation pipelines. In this study, we take the first steps toward this goal by collecting, studying, and making available\footnote{The corpus will be available at \url{https://voices.uchicago.edu/p2r/bpc-cpd-corpus/}
under terms of use in line with US federal standards for human subjects research. Researchers seeking access to the data 
will be asked to agree to the terms of use
 and to describe their intended research use case. Each use case will be manually reviewed for ethical considerations before access is granted.}
 a corpus of BPC in Chicago, one of the largest police forces in the United States.  Specifically, we investigate automatic speech recognition (ASR) model performance on a corpus of two-way radio transmissions between policing professionals in the city of Chicago to characterize challenges associated with this domain. Our main contributions are (1) the corpus itself and (2) our evaluation of ASR model performance, which provides a reference point for the current capabilities and limits of such models on this challenging domain. 

\section{Background}
\subsection{Domain characteristics of police radio speech}
\vspace{-.05in}
Police have used two-way radio transmissions to coordinate their activity since the 1930s \cite{citizens_police_committee_chicago_1931, poli_development_1942}. 
A large number of these transmissions are short utterances involving a handshaking-like process where individuals confirm each other's identity before transmitting information. These utterances often contain a unit number identifying the police beat (i.e., geographic region) the unit is assigned to patrol, and are therefore important for speaker identification and localization (e.g., ``FOURTEEN TWELVE" is a call sign indicating the unit is patrolling beat number 1412). Beyond their use in call signs, numbers are an essential part of this domain since they are used for street addresses, ages of individuals encountered, and other information (e.g., ``TEN FOUR" to acknowledge a transmission). 
\textcolor{black}{Representative utterances from our BPC corpus include ``WE GOT FIRST FLOOR I THINK IT'S APARTMENT NUMBER F ONE OR L ONE" and ``TWO TWENTY TWO ROBERT WE ARE ON OUR TRAFFIC CRASH \textlangle UNINTELLIGIBLE\textrangle," where "\textlangle UNINTELLIGIBLE\textrangle" indicates a segment that the annotator could not understand (see Section~\ref{sec:data}).}  Finally, radio transmissions using the same frequency can interfere with each other. In practice, this means speakers engage in strict turn-taking.

Another feature of radio policing, like other police data~\cite{prabhakaran2018detecting}, is the sharing of sensitive information about individuals, such as name and address. However, like other forms of policing data, BPC are often publicly accessible and may be observed and/or collected for research according to applicable local laws.  Therefore, our data can be released to other researchers under applicable guidelines, and our research methods should be reproducible by other researchers aiming to analyze BPC data in other regions.  In addition to the data, we will make our materials, such as annotation guidelines and data processing pipelines, available to the research community.  

\subsection{Speech recognition for naturalistic audio}
Recently, the research community has shifted from using curated benchmark corpora, such as LibriSpeech~\cite{Panayotov2015LibrispeechAA}, to training on as much accessible speech data as possible \cite{radford2022whisper,peng2023reproducing,kang_libriheavy_2024}. Using large amounts of diverse data can help create general-purpose ASR systems that can be used in a zero-shot manner (i.e.,~without training on domain-specific data). However, domain differences can still degrade performance. For example, the Whisper large-v2 speech recognizer~\cite{radford2022whisper} degrades from 2.7\% word error rate (WER) on LibriSpeech to 25.5\% WER on data from the CHiME6 noisy speech challenge~\cite{watanabe2020chime} and 36.4\% WER on multi-party meeting speech from the AMI corpus~\cite{renals2007recognition}.  On clean but accented speech, Whisper degrades to 19.7\% WER~\cite{sanabria2023edinburgh}. 
Here, we focus on the domain of police radio communications, where unique challenges such as background noise, specialized terminology, and short utterances further compound the challenges of speech recognition in real-world conditions.  In some ways this domain is similar to the air traffic control (ATC) domain, and there has been work on ASR for ATC~\cite{pellegrini_airbus_2019,zuluaga2023does}. 
 However, word error rates for ATC corpora tend to be much lower than for BPC (see Section~\ref{sec:results}), likely reflecting differences in audio quality despite the domain similarities. 

\section{Data}
\vspace{-.1in}

Our corpus, BPC-CPD, is composed of BPC radio transmissions involving the Chicago Police Department (CPD). BPC-CPD 
contains \textcolor{black}{$62,080$} utterances (\textcolor{black}{$46.2$} hours of transcribed speech) labeled by a combination of researchers at the University of Chicago 
(\textcolor{black}{42.5\%}) and a professional transcription service (\textcolor{black}{57.5\%}). The  professional service transcribed 510 minutes of ``raw" audio (i.e., including silence between transmissions) so that these could be aligned with internal transcribers' annotations for validation. The remaining professionally transcribed audio was sampled from all of the speech in the BPC recordings, as identified by a voice activity detection model. 
BPC-CPD includes data from 11 of Chicago's 13 dispatch zones, distinct non-overlapping areas of the city each assigned one dispatcher to coordinate police activity in that area; data for the remaining two dispatch zones was not available. 

\begin{figure}[t]
  \centering
  \includegraphics[scale=0.55]{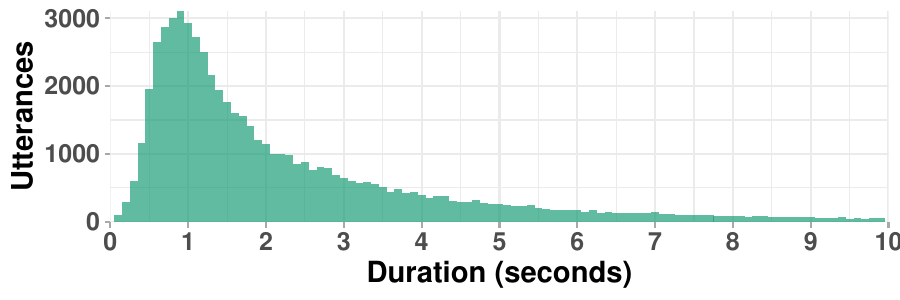}
  \caption{Duration distribution of utterances in our training set, for utterances of duration up to 10 seconds (96.5\% of the train set utterances).}
  \label{fig:hist_train}
  \vspace{-15pt}
\end{figure}

 The training, development, and test sets consist of \textcolor{black}{$44,664$} utterances (\textcolor{black}{33.0} hours), \textcolor{black}{$8,714$} utterances (\textcolor{black}{6.6} hours), and \textcolor{black}{$8,702$} utterances (\textcolor{black}{6.6} hours), respectively. Train (80\%), dev (10\%), and test (10\%) splits were created through stratified sampling of transcribed utterances. To ensure even sampling of BPC across dispatch zones and day/night hours, we created 22 strata (11 zones $\times$ 2 levels), since zone captures spatial variation in reported crimes and policing personnel, while the timing of police stops (before/after sunset) impacts police behavior \cite{pierson_large-scale_2020}. The utterances from two zones (5 and 6) were evenly split into dev/test sets, but excluded from train in order to assess robustness to unseen contexts. BPC associated with the remaining 9 zones are split between train, dev, and test as described above. Figure \ref{fig:hist_train} shows the utterance length distribution for the train split, which is essentially identical to the dev and test distributions.

\vspace{-.05in}
\subsection{Data collection and annotation process}
\label{sec:data}

BPC-CPD data were obtained from a publicly accessible web site (https://broadcastify.com) that maintains a rolling 6-month archive of recordings available under a CC-BY-3.0 license. Approximately 12 months (\textasciitilde80k hours) of BPC recordings were downloaded, of which $46.2$ hours have been labeled thus far. The BPC-CPD corpus contains transcriptions for these 46.2 hours, including transcripts from multiple annotators for the same audio when available. Here, we use a deduplicated version of the corpus for development and testing, selecting the transcript produced by the annotator with the lowest median WER (see Section~\ref{sec:iaa}). 

Annotation guidelines were developed based on pilot transcription efforts by researchers at the University of Chicago, with researchers collectively discussing and addressing annotation challenges as part of protocol development. Internal annotators were provided 30-minute audio files and asked to identify and transcribe any speech found in the file, treating each radio transmission as a separate utterance and spelling out numbers as spoken. Annotators were instructed not to download audio, and annotation took place in a secure compute environment: Access to audio was restricted to either secure on-site computers or secure remote desktop applications. Annotators were made aware of the sensitive nature of BPC and advised to pause work if they experienced emotional distress given its content. 

If annotators were uncertain about a transcription, they used two types of in-line annotation: (a) enclose words they are 50\%-90\% sure are correct in square brackets (e.g., ``[GOOD JOB]"); (b) indicate a segment of speech is unintelligible (i.e., $<$50\% sure they correctly understood it) using ``$\langle x \rangle$" (henceforth denoted ``\textlangle \textcolor{black}{UNINTELLIGIBLE}\textrangle" in our analysis). This annotation protocol was provided to the professional transcribers, along with a list of Chicago street names and common CPD jargon. 

Before further analysis and ASR experiments, we post-processed the annotations to correct typos and normalize the text for ASR.

\subsection{Post-processing steps}\label{sec:prepro}

For purposes of ASR training and evaluation (and the human WER evaluation below), all square brackets indicating transcriber uncertainty were removed, all hyphens were replaced with whitespace, single quotes and backticks were converted to apostrophes, and all punctuation except for apostrophes was removed. In addition, certain annotator formatting errors were also manually corrected. Finally, annotation errors like misspellings were identified and, where possible, corrected following manual review. 

This process used a whitelist of tokens considered valid based on a combination of NLTK's English words and Chicago street names~\cite{bird2009natural}. 
The manual review process 
leveraged both context and domain knowledge about policing in Chicago to correct tokens when (a) the error is clearly a typo, (b) the intended word or words are clear, given context, or (c) domain knowledge suggested a specific correction (e.g., ``KEDZIE" is a street name that has many misspellings). If significant ambiguity persisted despite context, then no change was made.

Numbers were converted to their spoken word representation (e.g., ``SIX" instead of ``6"). 
First, as many numbers as possible (where there was no ambiguity) were corrected via the review process described above. For example, the number ``$720$'' would be corrected to ``SEVEN TWENTY" since this is its conventional spoken word representation in BPC. 
Second, for consistency the NeMo text normalizer was applied to remaining numbers whose spoken word representation was unclear or ambiguous~\cite{zhang21ja_interspeech, bakhturina22_interspeech}. Finally, all text was upper-cased. 

In the case of utterances that were transcribed by multiple annotators, for the training set the ground-truth transcript was selected randomly.
For the development and test set, the selection was done using annotator performance; specifically, the transcript from the annotator with the lowest median WER against other annotators was chosen. 
All post-processing steps described in this section are also applied to ASR model outputs for comparability. 

The released BPC-CPD corpus will include two versions of all transcriptions: (1) raw text as entered by annotators; (2) post-processed text, as used here, but including brackets to indicate annotator uncertainty. We do not make use of these in-line annotations in our current study, but we include them in the released corpus for use by others for robustness testing or other research uses. 
We will also release the complete details of our annotation and post-processing pipeline, to facilitate replication of our pipeline for BPC data collection in other regions. 

\subsection{Inter-annotator agreement}
\label{sec:iaa}


We report inter-annotator agreement using the WER between annotators for audio transcribed by multiple annotators. Since each annotator may have a different set of utterance boundaries, for each pair of annotators we identify audio files transcribed by both annotators and then concatenate all utterances from each file. Since we lack ground truth, each annotator's transcript is treated first as reference and then as hypothesis. Across 23 annotators who contributed to BPC-CPD, we 
have \textcolor{black}{468} triplets of the form (transcription from annotator $i$, transcription from annotator $j$, audio file), where WER for the triplet is calculated using the transcription from the first annotator as reference. 

According to this evaluation, the professional annotators have a median WER of 25.9\% (when treating all other annotators as reference), while the remaining annotators have a median WER of 28.9\%. These WERs provide a rough measure of human-level performance on our task.  This measure is not directly comparable to the performance of our ASR systems, which are given a single utterance at a time without context, rather than the much longer audio that human transcribers were given.  We also note that the true human WER is likely somewhat lower, since there is some noise introduced in the transcriptions through differences in spelling that our post-processing (see Section~\ref{sec:prepro}) may have missed.

Nevertheless, the relatively high annotator WERs suggest that BPC audio is challenging even for human listeners.  We also note that roughly \textcolor{black}{13\%} of the utterances in our corpus contain at least one audio segment marked unintelligible by a transcriber.  This finding is consistent with interviews conducted with sworn officers about both BPC and interactions with youth, during which many asked for BPC audio clips to be repeated due to difficulty understanding what was spoken. While the most common confusions are the expected ones between short function words, the single most commonly disagreed-upon word is ``\textlangle unintelligible\textrangle".
\begin{table*}[t]
    \caption{WER (\%) for ASR models on the dev and test sets, as well as the substitution (S), deletion (D), and insertion (I) rates (\%) for the dev set.  We evaluate only the best models in each category on the test set, to avoid overfitting to the test set. 
    [$^*$These models have fewer tuned parameters than the fused models due to projecting features to a lower dimensional space prior to inputting them to the encoder.]}
    \centering
    \normalsize
    \resizebox{\textwidth}{2.3in}{
    \begin{tabular}{l| l |r r c |c r r r |c}
    \toprule
        Type & Model  & \makecell{\# \\ Parameters} & \makecell{\# Tuned\\ Parameters} & \makecell{Language\\ Model?} & WER$_{dev}$ & S$_{dev}$ & D$_{dev}$ & I$_{dev}$ & WER$_{test}$ \\ \midrule \midrule
            \cmidrule{2-10}
        Whisper & Whisper large-v2  & 1.55B & 0 & No & 57.4 & 25.1 & 15.8 & 16.5 \\ 
        & Whisper large-v3  & 1.55B & 0 & No & 51.4 & 26.2 & 11.2 & 14.0 & 50.8\\ \midrule \midrule
        NeMo & Conformer CTC &  120M & 0  & No  & 51.5  & 27.7 & 19.4 & 4.4 \\
        &   &  120M & 120M  & No  & 27.7 & 13.5 & 10.5 & 3.6 \\
        &  &  616M & 0  & No  & 53.8 &  29.7 & 19.7 & 4.4 \\ 
        &   &  616M & 616M  & No  & 27.0  & 13.8 & 9.2 & 4.0 & \textbf{27.3} \\
       &  &  1.2B & 0  & No  & 53.1 & 27.3 & 21.7 & 4.1 \\ \cmidrule{2-10} 
       & Conformer parakeet CTC &  616M & 0  & No  & 53.8  & 28.6 & 21.1 & 4.1 \\ 
       &  &  1.2B & 0  & No  & 51.6 &  29.5 & 16.6 & 5.5 \\ \cmidrule{2-10} 
       & Conformer parakeet Transducer &  616B & 0  & No  & 54.9 & 18.6 & 33.5 & 2.8 \\ 
       &  &  1.2B & 0  & No  & 50.8 & 19.9 & 28.1 & 2.8 \\ \cmidrule{2-10} 
      &  Conformer parakeet TDT &  1.2B & 0  & No  & 48.2  & 22.2 & 21.9 & 4.1 &  47.5 \\ \midrule \midrule
      CTC-AED &  FBANK-Conformer & 41M & 41M & No & 51.9  & 33.6 & 9.0 & 9.3 \\ 
      &   & 41M & 41M & Yes & 51.3 & 32.1 & 10.5 & 8.7 \\ \cmidrule{2-10} 
      &  HuBERT-Conformer & 422M & 106M   & No & 52.3 &  33.9 & 9.0 & 9.4 \\ 
      &   & 422M & 106M & Yes & 51.1 &   32.0 & 9.6 & 9.5 \\  \cmidrule{2-10} 
      &  WavLM-Conformer & 421M & 106M & No & 42.3  & 25.9 & 6.5 & 9.8 \\ 
      &   & 421M & 106M & Yes & 40.6 &  23.9 & 7.0 & 9.7 \\ \cmidrule{2-10} 
      &  HuBERT+WavLM-Conformer & 683M & 52M$^*$ & No & 40.1 & 24.4  & 6.8 & 8.8 & 39.2 \\
      &   & 638M & 52M$^*$ & Yes & 39.9 & 23.5 & 7.4 & 8.9 & 39.4  \\ \cmidrule{2-10} %
      &  FBANK-Transformer & 7M & 7M & No & 60.3  & 38.3 & 11.9 & 10.1 \\ 
      &  HuBERT-Transformer & 340M & 23M  & No & 58.7  & 36.4 & 13.8 & 8.4 \\ 
      &  WavLM-Tranformer & 339M & 23M & No & 44.0 &  26.0 & 9.8 & 8.1 \\ 
      &  HuBERT+WavLM-Transformer & 642M & 10M$^*$ & No & 42.2  & 25.4 & 8.1 & 8.7 \\ 
      \bottomrule
    \end{tabular}}
    \label{tab:wer}
\end{table*}

\section{Experimental setup}

\subsection{Off-the-shelf ASR models}
We start by evaluating how well off-the-shelf ASR models from the Whisper and NeMo families perform on the BPC domain without any fine-tuning. For the Whisper family, we use Whisper large-v2~\cite{radford2022whisper} and Whisper large-v3 \cite{OpenAI_Whisper}, which were trained on 680k hours and \textasciitilde5M hours, respectively. They share the same architecture: 32 transformer layers with 20 attention heads per layer for both audio encoder and text decoder. 

For the NeMo models, we test token-and-duration transducer (TDT)-based~\cite{xu2023efficient}, transducer-based~\cite{graves2012sequence}, and connectionist temporal classification (CTC)~\cite{graves2006connectionist} models, all of which were trained on 64k hours of English data. The architecture of these models is based on the fast conformer with linear attention~\cite{rekesh2023fast}. NeMo models have 8 attention heads per layer, and come in several sizes: large (L, $17$ layers and 120M parameters), extra large (XL, $24$ layers and 616M parameters), and extra-extra large (XXL, $42$ layers and 1.2B parameters).

\subsection{Fine-tuned ASR models}
We fine-tune the pre-trained NeMo models to measure the extent to which optimizing parameters for the domain helps.
For fine-tuning experiments, we choose the CTC-based NeMo models because they have a good performance vs.~speed tradeoff.\footnote{Based on the open ASR leaderboard: \url{https://huggingface.co/spaces/hf-audio/open_asr_leaderboard}.}
We keep the original tokenizer when fine-tuning the models. We use the AdamW~\cite{loshchilov2017decoupled} optimizer with a weight decay of $0.001$. We tune the learning rate over $2 \times 10^{-5}$, $5 \times 10^{-5}$, and $10^{-4}$, and select the best models based on WER on the dev set. We use cosine learning rate annealing with $5000$ warm-up steps.

\begin{table*}[t]
\label{t:model_conf}
\caption{Top 10 most common ASR model confusions. 
``X $\longrightarrow$ Y"
 indicates that the ground-truth word ``X" was misrecognized as the hypotheized word ``Y".}
    \centering
    \begin{tabular}{ccr|ccr|ccr|ccr|ccr}
    \multicolumn{3}{c|}{custom model, no LM}            &  \multicolumn{3}{c|}{custom model, LM}                  &  \multicolumn{3}{c|}{NeMo, no fine-tune} &  \multicolumn{3}{c}{NeMo, fine-tune}  &  \multicolumn{3}{c}{Whisper, large-v3}\\ \hline
 the & $\longrightarrow$ & a                            & the  & $\longrightarrow$ & \textlangle unint\textrangle & four   & $\longrightarrow$ & for     & one's & $\longrightarrow$ & one      &  two & $\longrightarrow$ & twenty   \\
   a & $\longrightarrow$ & the                          & four & $\longrightarrow$ & \textlangle unint\textrangle & two    & $\longrightarrow$ & to      & the   & $\longrightarrow$ & a        &  three  & $\longrightarrow$  & thirty   \\
  oh & $\longrightarrow$ & zero                         & on   & $\longrightarrow$ & \textlangle unint\textrangle & a      & $\longrightarrow$ & the     & a     & $\longrightarrow$ & the      &  gonna  &  $\longrightarrow$ & to \\
four & $\longrightarrow$ & \textlangle unint\textrangle & a    & $\longrightarrow$ & the                          & gonna  & $\longrightarrow$ & to      & zero  & $\longrightarrow$ & oh       &  zero & $\longrightarrow$  & and  \\
 the & $\longrightarrow$ & \textlangle unint\textrangle & a    & $\longrightarrow$ & \textlangle unint\textrangle & the    & $\longrightarrow$ & a       & ah    & $\longrightarrow$ & uh       &  one  & $\longrightarrow$  & and \\
  in & $\longrightarrow$ & and                          & and  & $\longrightarrow$ & \textlangle unint\textrangle & two    & $\longrightarrow$ & you     & two   & $\longrightarrow$ & six      &  six &  $\longrightarrow$ & sixty   \\
  on & $\longrightarrow$ & \textlangle unint\textrangle & the  & $\longrightarrow$ & a                            & and    & $\longrightarrow$ & in      & \textlangle unint\textrangle & $\longrightarrow$ & uh & five  & $\longrightarrow$  & fifty \\
 and & $\longrightarrow$ & \textlangle unint\textrangle & two  & $\longrightarrow$ & \textlangle unint\textrangle & one's  & $\longrightarrow$ & one     & seven & $\longrightarrow$ & eleven   & four  & $\longrightarrow$  & forty  \\
   a & $\longrightarrow$ & \textlangle unint\textrangle & you  & $\longrightarrow$ & \textlangle unint\textrangle & uh     & $\longrightarrow$ & ah      & three & $\longrightarrow$ & two      & four & $\longrightarrow$  & you  \\
 two & $\longrightarrow$ & \textlangle unint\textrangle & to   & $\longrightarrow$ & \textlangle unint\textrangle & robert & $\longrightarrow$ & roberts & thirty& $\longrightarrow$ & three    &  a & $\longrightarrow$  & the  \\
    \label{tab:confusions}
\end{tabular}
\vspace{-20pt}
\end{table*}

\subsection{Customized E2E ASR models}
We implement our customized models using ESPNet~\cite{watanabe2018espnet} and S3PRL (for self-supervised learning (SSL) model support)~\cite{Yang2021SUPERBSP}. These models test the value of adapting the parameters, token vocabulary and language model to the domain. We use hybrid CTC - attention-based encoder decoder (CTC-AED) models due to their good performance across many tasks \cite{Watanabe2017HybridCA}. We explore both transformer and conformer encoder architectures. Our transformer encoders have 12 layers with 8 attention heads per layer and an output dimension of 128. The conformer encoders consist of 12 layers with 4 attention heads per layer and output dimension of 256. For all experiments, the decoder is a 6-layer transformer. These hyperparameters were chosen based on tuning experiments on the validation set. We also integrate an RNN (specifically LSTM) language model (LM) \cite{Mikolov2010RecurrentNN} trained on BPC-CPD, which has a perplexity of 11.6 on the dev set. 

For each model, we test four feature extractors: log Mel-filterbank features (FBANK), HuBERT Large (300M)~\cite{Hsu2021HuBERTSS}, WavLM Large (300M)~\cite{Chen2021WavLMLS}, and a feature fusion model, combining representations learned by HuBERT and WavLM. HuBERT and WavLM were chosen based on their consistently good performance in various benchmarks~\cite{Yang2021SUPERBSP, Chen2021WavLMLS}. 
As is common in recent benchmarks~\cite{Yang2021SUPERBSP}, we use a weighted sum of frozen self-supervised model layers as input, with the weights learned during fine-tuning. For the feature fusion experiment, we follow the approach of \cite{Berrebbi2022CombiningSA, Chen2022FeaRLESSFR, effuse}.

\section{Results and analysis}
\label{sec:results}

\subsection{ASR performance}
Table~\ref{tab:wer} shows our main ASR results.  Among the pre-trained ASR models, Whisper large-v3 outperforms Whisper large-v2 by a large margin, but is slightly worse than NeMo models in general. For NeMo models, the larger models sometimes but not always improve WER, suggesting that scaling up the model does not necessarily address domain differences. In terms of model type (CTC, transducer, and TDT), the transducer and TDT models have better performance in general. We leave fine-tuning these two models as future work.



After fine-tuning the NeMo Fast-Conformer CTC model on BPC data, we see a dramatic improvement in WER, suggesting that fine-tuning can bridge much of the domain difference between the pre-trained model and the police radio domain. See Section~\ref{sec:error} for further analysis.

For the customized E2E models, conformer based models outperform their transformer counterparts. Interestingly, the performance of the FBANK-conformer model is similar to that of many of the pre-trained models in the NeMo and Whisper families, suggesting that off-the-shelf ASR can be replaced with smaller models if some domain data is available for training. The language model provides a small improvement to most of the ASR models, but the best-performing CTC-AED model shows the least effect from the use of the language model.  Overall, the usefulness of a language model for this domain is unclear, and merits further study with different language models. 

Performance on the ``unseen" Zone 5 (WER$_{dev}=30\%$, WER$_{test}=31\%$) and Zone 6 (WER$_{dev}$ =WER$_{test}=30\%$) 
is almost as good as on the previously seen zones.
This robustness check is especially useful since dispatch zones differ in terms of speakers, locations, and radio instrumentation (e.g., location/configuration of antenna).

\begin{figure}[h!]
  \centering
  \includegraphics[scale=0.55]{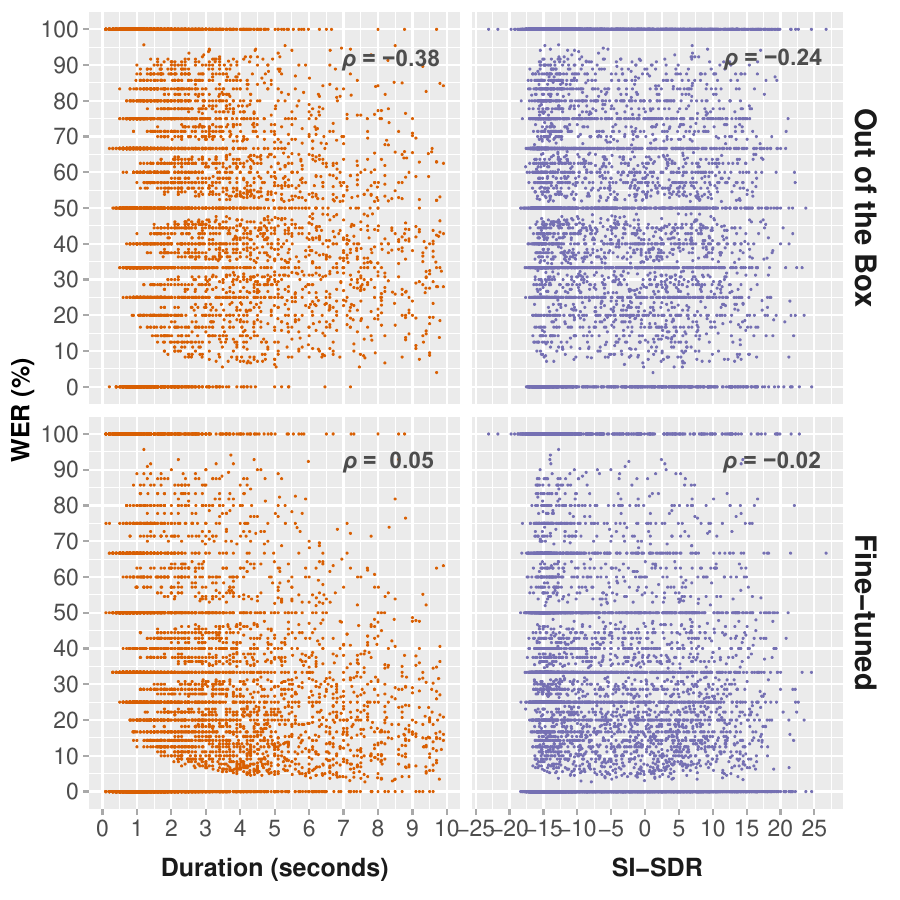}
  \caption{WER vs.~audio quality (SI-SDR) and utterance duration for NeMo FastConformer CTC (616M) on the dev set.}
  \label{fig:wer_dur_sisdr}
  \vspace{-15pt}
\end{figure}

\vspace{-.1in}
\subsection{Error analysis}\label{sec:error}
\subsubsection{Common errors}

Table \ref{tab:confusions} shows the most common ASR model confusions for five of our top-performing models.  The off-the-shelf models have the most trouble with domain-specific words like digits and ``robert" (code word for the night shift).  The custom model, as well as the NeMo model after fine-tuning, do not suffer from as many of these domain-specific confusions, instead having more common stop words (``a", ``the") among their top errors, as do human annotators. 
In addition, neither NeMo nor Whisper have the ``unintelligible" token in their vocabularies, so they cannot label a speech segment as unintelligible.  On the other hand, the custom models often predict the unintelligible token incorrectly, and it features prominently among their most common errors. 


\subsubsection{Effect of utterance length and audio quality on errors}
Figure~\ref{fig:wer_dur_sisdr} shows scatterplots of the per-utterance WERs of NeMo FastConformer CTC (616M) for utterances of different lengths and estimated audio quality levels. We include results for both off-the-shelf and fine-tuned models, along with the Spearman correlation coefficient between WER and duration/audio quality.
Audio quality is estimated via a neural quality estimator~\cite{kumar2023torchaudio} trained to predict SI-SDR~\cite{le2019sdr}. Utterance duration exhibits a weak inverse relationship with WER, and fine-tuning appears to mitigate this relationship (i.e., the correlation between WER and duration is decreased). 
Audio quality also has a weak inverse relationship with WER, which also flattens after fine-tuning.  
Overall, fine-tuning has the intended effect of both reducing WER (globally and among highly prevalent short utterances) \textit{and} reducing the effect of audio quality on WER. 

\section{Conclusions and future work}
Our study of automatic speech recognition on the BPC domain, which serves a functional role in police communication, is the first step toward better automated analysis of this form of police communication. Our findings provide a baseline assessment of how current speech recognition models perform in this challenging domain.  We provide access to the BPC-CPD corpus so that others can develop and test novel solutions to the challenges associated with BPC, and to enable future analysis of policing practices through BPC.

Perhaps as expected given the domain, we have found that existing pre-trained models perform poorly (WER $\ge$ \textasciitilde50\%). 
Fine-tuning is effective, with our best performing model achieving a WER of 27.3\%, which is roughly in the same range as inter-annotator agreement on our data. The relationship between WER and estimated audio quality (SI-SDR) is weak, suggesting that explicit denoising methods are unlikely to be of great help.  We also found that fine-tuning mitigates some of the impact of noise. 

One limitation of this work is the location-specific source of BPC data. BPC from other police systems may exhibit different properties with fewer, different, or more challenges.  However, there are also many shared properties across police radio domains in different locations, so we believe that our experience can help guide researchers studying this data domain elsewhere.

Considering the difficulty of annotation in this domain, future work includes approaches for effectively using unlabeled data in the domain, and analysis of how the quantity and quality of labeled data impacts performance. Further inter-annotator analysis may also allow us to assess, for example, whether audio marked unintelligible is linked to miscommunication.
Considering the prolific use of BPC in police departments around the world, this kind of exploration is needed to understand how BPC shapes officer decision-making.

\section{Acknowledgments}
Research supported by National Institute of Minority Health and Health Disparities of the National Institutes of Health under award number R01MD015064. This research used resources of the Argonne Leadership Computing Facility, a U.S. Department of Energy (DOE) Office of Science user facility at Argonne National Laboratory and is based on research supported by the U.S. DOE Office of Science-Advanced Scientific Computing Research Program, under Contract No. DE-AC02-06CH11357. 

\section{References}
\printbibliography

\end{document}